\documentclass[conference]{IEEEtran}
\IEEEoverridecommandlockouts
\usepackage{cite}
\usepackage{amsmath,amssymb,amsfonts}
\usepackage{algorithmic}
\usepackage{graphicx}
\usepackage{textcomp}
\usepackage{xcolor}
\usepackage{hyperref}
\usepackage{tikz}
\usepackage{orcidlink}
\usepackage{tikz}
\usetikzlibrary{shapes.geometric, arrows.meta, positioning, calc}
\usepackage{xcolor}
\hypersetup{pdfborder={0 0 0}}
\usepackage{gensymb}

\usetikzlibrary{shapes.geometric, arrows, positioning, calc, fit}
\def\BibTeX{{\rm B\kern-.05em{\sc i\kern-.025em b}\kern-.08em
		T\kern-.1667em\lower.7ex\hbox{E}\kern-.125emX}}

\definecolor{lightblue}{RGB}{66,133,244}

\newcommand{\preprintbanner}{%
	\begin{tikzpicture}[remember picture, overlay]
		\node[rectangle, text=red!80!white, inner sep=8pt,
		anchor=north, text width=\paperwidth, align=center] 
		at ([yshift=-0.5cm]current page.north) 
		{Author pre-print. Publication accepted for \href{https://conf.researchr.org/home/edtconf-2025}{\textcolor{lightblue}{EDTconf 2025}}.};
	\end{tikzpicture}%
}

\newcommand{\preprintbannerbottom}{%
	\begin{tikzpicture}[remember picture, overlay]
		\node[rectangle, text=red!80!white, inner sep=8pt,
		anchor=south, text width=\textwidth, align=center] 
		at ([yshift=0.5cm]current page.south) 
		{\scriptsize
			© 2025 IEEE. Personal use of this material is permitted. Permission from IEEE must be obtained for all other uses, in any current or future media, including reprinting/republishing this material for advertising or promotional purposes, creating new collective works, for resale or redistribution to servers or lists, or reuse of any copyrighted component of this work in other works.
			
			\vspace{0.3em}};
	\end{tikzpicture}%
}

\begin{document}
	
	\title{Engineering a Digital Twin for the Monitoring and Control of Beer Fermentation Sampling}
	
	\author{
		\IEEEauthorblockN{Pierre-Emmanuel Goffi\orcidlink{0009-0003-3420-6955}}
		\IEEEauthorblockA{\textit{Dept. of Computer and Software Eng.} \\
			\textit{Polytechnique Montréal}\\
			Montréal, Canada \\
			pierre-emmanuel.goffi@polymtl.ca}
		\and
		\IEEEauthorblockN{Raphaël Tremblay\orcidlink{0009-0009-7590-9124}}
		\IEEEauthorblockA{\textit{Dept. of Computer and Software Eng.} \\
			\textit{Polytechnique Montréal}\\
			Montréal, Canada \\
			raphael-1.tremblay@polymtl.ca}
		\and
		\IEEEauthorblockN{Bentley Oakes\orcidlink{0000-0001-7558-1434}}
		\IEEEauthorblockA{\textit{Dept. of Computer and Software Eng.} \\
			\textit{Polytechnique Montréal}\\
			Montréal, Canada \\
			bentley.oakes@polymtl.ca}
	}
	
	\maketitle
	
	\preprintbanner
	\preprintbannerbottom
	
	\begin{abstract}
Successfully engineering interactive industrial DTs is a complex task, especially when implementing services beyond passive monitoring. We present here an experience report on engineering a safety-critical digital twin (DT) for beer fermentation monitoring, which provides continual sampling and reduces manual sampling time by 91\%. We document our systematic methodology and practical solutions for implementing bidirectional DTs in industrial environments. This includes our three-phase engineering approach that transforms a passive monitoring system into an interactive Type 2 DT with real-time control capabilities for pressurized systems operating at seven bar. We contribute details of multi-layered safety protocols, hardware-software integration strategies across Arduino controllers and Unity visualization, and real-time synchronization solutions. We document specific engineering challenges and solutions spanning interdisciplinary integration, demonstrating how our use of the constellation reporting framework facilitates cross-domain collaboration. Key findings include the critical importance of safety-first design, simulation-driven development, and progressive implementation strategies. Our work thus provides actionable guidance for practitioners developing DTs requiring bidirectional control in safety-critical applications.
\end{abstract}

	\section{Introduction}

Beer brewing stands as one of humanity’s oldest biotechnological processes, dating back thousands of years. Since its ancient origins, brewing has evolved into a sophisticated operation balancing craftsmanship with cutting-edge technology.

This project concerns the beer fermentation phase, which is the most critical and complex stage in beer production where yeast transforms sugars into alcohol, CO$_2$, and flavor compounds. This process typically occurs over days to weeks, depending on the beer style, with each phase requiring specific environmental
conditions for optimal outcomes. The precise management of this process thus directly impacts product quality, consistency, and production efficiency~\cite{Josey2019current}.

 This fermentation process typically requires weeks of careful monitoring in temperature-controlled tanks. Traditional quality control relies on master brewers manually extracting samples to measure parameters such as specific gravity, dissolved oxygen, pH, and conductivity. This manual approach presents several challenges: it is time-consuming, wastes product, introduces contamination risks, and provides only periodic snapshots rather than continuous monitoring.

 \paragraph*{Physical and Digital Twins for Fermentation}
 At our university Polytechnique Montr\'{e}al, the technical society Polybroue\footnote{\url{https://www.facebook.com/p/Polybroue-100063519554992/}} has developed a fermentation monitoring prototype capable of conducting mechanical tests. However, this system previously lacked software automation to realize its full potential as a research and process control tool.
 
Our digital twin (DT) system addresses these limitations by automating measurement and providing real-time visibility into fermentation dynamics. This paper thus reports on the \textit{engineering of this DT solution for beer fermentation monitoring}, creating a dynamic relationship between physical brewing equipment and its digital representation. This DT was created as a project during the \textit{Digital Twin Engineering} graduate course at Polytechnique Montr\'{e}al, taught by the last author\footnote{\url{https://bentleyjoakes.github.io/dte_course/}}.

The implemented solution comprises two primary components: 1) a physical twin utilizing an Arduino-based control system that manages a specialized pressurized measurement chamber, and 2) a DT hosted on a Linux server for real-time visualization, data storage, and limited control capabilities.

\begin{figure}[ht!]
    \centering
    \includegraphics[width=0.70\linewidth]{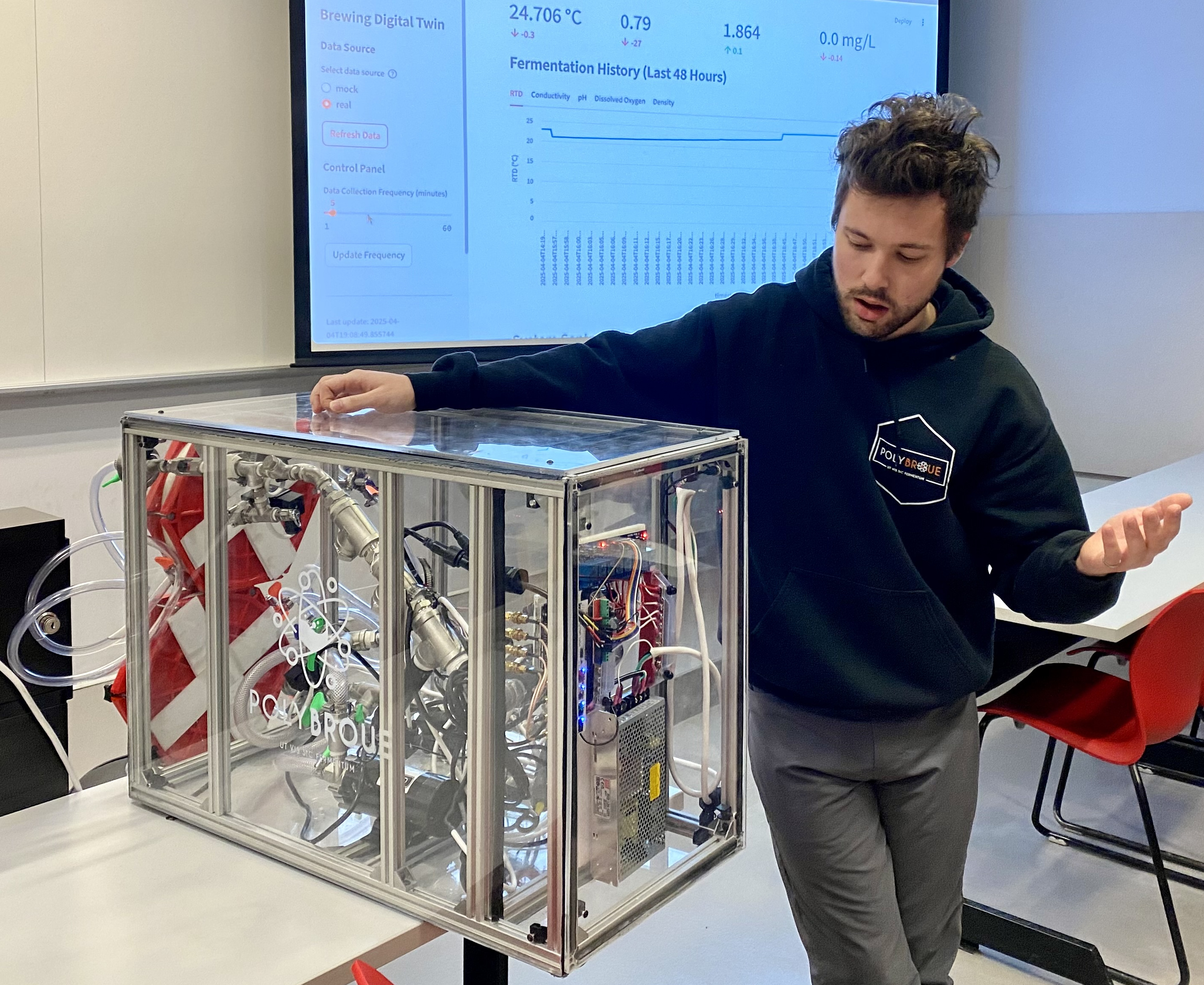}
    \caption{Presenting the fermentation monitoring tank and dashboard.}
    \label{fig:PT}
\end{figure}

The physical system (seen in Figure~\ref{fig:PT}) circulates fermentation contents through a sampling chamber. There, it pressurizes the contents to seven bar\footnote{7 bar is about 3 times standard tire pressure or 70 meters of water pressure.} using nitrogen, and measures critical parameters including dissolved O$_2$, pH, conductivity, and temperature, while an electronic hydrometer continuously monitors density data directly from the fermentation tank.

The DT is a `true digital twin' (by the classification of Kritzinger \textit{et al.}~\cite{Kritzinger2018}) through implementation of bidirectional communication protocols that enable the digital representation to adjust non-critical parameters while maintaining robust safety constraints. The system incorporates sophisticated control mechanisms for critical decompression phases with safety protocols that prevent equipment damage and operational hazards. A comprehensive Unity-based visualization system provides brewers with intuitive 3D representation of fermentation processes and equipment status, supported by a database solution that tracks parameter evolution over time.

\begin{figure*}[tbh]
\centering
\includegraphics[width=0.32\linewidth]{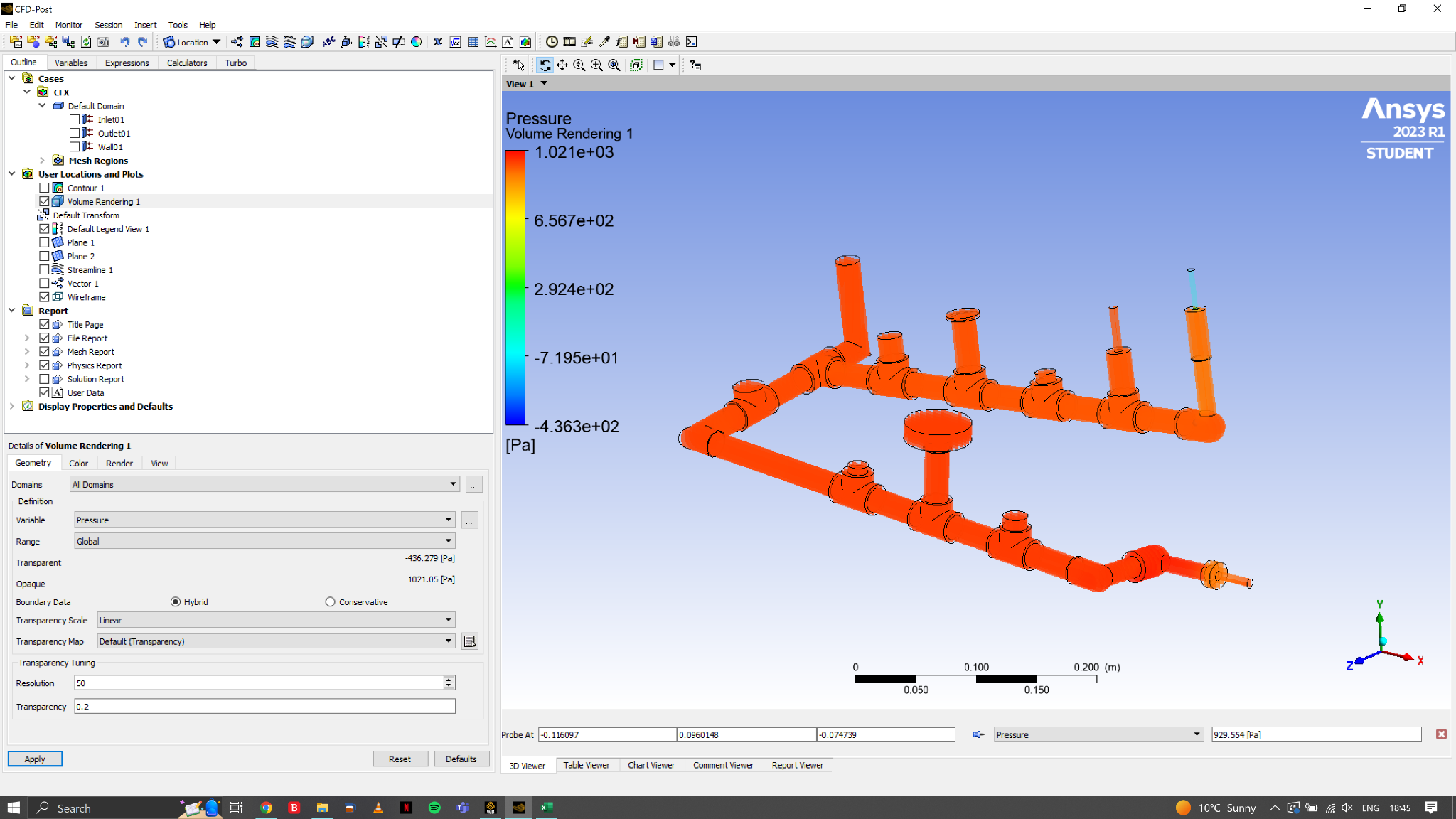}
\includegraphics[width=0.32\linewidth]{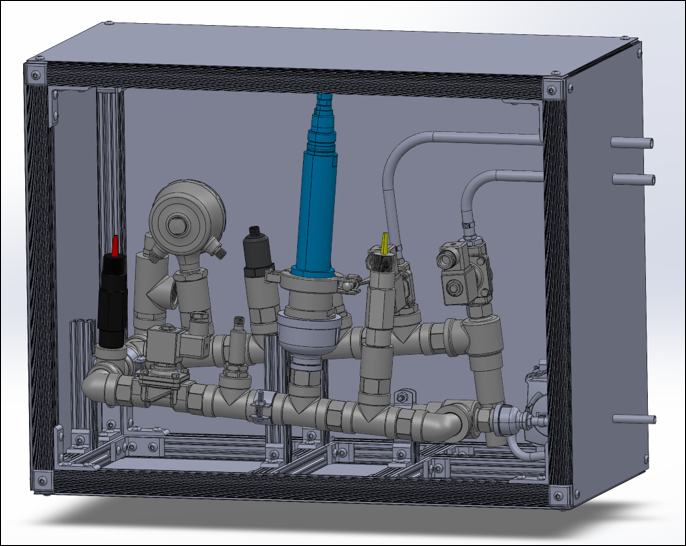}
\includegraphics[width=0.32\linewidth]{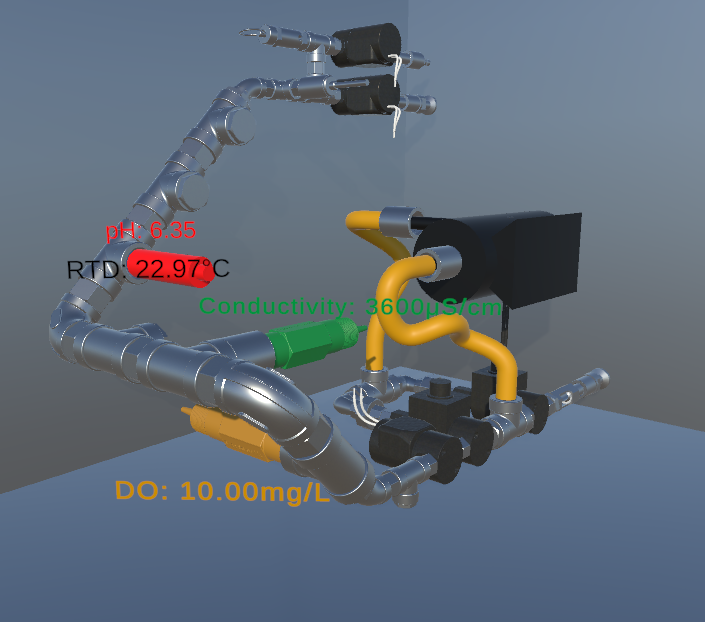}
\caption{DT Evolution: (a) Prototype sketch (Fall 2023), (b) CAD implementation (Fall 2024), and (c) Representation in Unity environment (Winter 2025).}
\label{fig:evolution}
\end{figure*}

This paper presents our insights on engineering a DT exemplar which has direct control on a complex physical system. We report on how this project addresses fundamental engineering challenges in brewing operations: inconsistent testing due to variable measurement techniques, limited visibility into process parameters during critical fermentation phases, and difficulty maintaining optimal conditions amid environmental fluctuations. The DT thus delivers continuous insights and control functions for consistent product quality, improved resource efficiency, and comprehensive process documentation. This work advances the state in digital twins by demonstrating practical implementation of bidirectional control in safety-critical industrial environments, addressing the gap between theoretical DT frameworks and operational deployment challenges. Unlike simulation-focused approaches, our contribution emphasizes real-time physical-digital synchronization with validated safety protocols for hazardous operations.

% \paragraph*{Contributions and Structure}

%  We explain the DT services and the architecture in Section~\ref{sec:services}. Following this, Section~\ref{sec:arch_and_impl} details the system architecture and implementation phases. 
% Section IV discusses the engineering process including development methodology, integration challenges, and validation procedures. Section V examines implementation challenges and solutions. Section VI presents results and discussion of system performance. Section VII provides related work context, and Section VIII concludes with summary and future directions.

	%\input{text/services}
	\section{Project Overview and Architecture}
\label{sec:arch_and_impl}

%This section describes the project's timeline, and the implementation of the physical fermentation monitor and its DT.

\paragraph{Project Timeline}

The engineering process employed a systematic three-phase approach transitioning from conceptual design to functional implementation as shown in Figure~\ref{fig:evolution}. The methodology emphasized iterative refinement with continuous validation at each development stage.

\textbf{Phase 1 - Conceptual Design/Prototyping (Fall 2023)}:

The initial engineering efforts focused on establishing system requirements through stakeholder analysis with brewing technicians and researchers. A functional decomposition identified critical subsystems including sampling mechanisms, pressurization control, sensor integration, and safety systems. The risk assessments evaluated potential failure modes, establishing safety requirements that informed later design decisions. This initial phase focused on non-invasive monitoring capabilities and safety considerations.

\textbf{Phase 2 - Mechanical/Electrical Integration (Fall 2024)}: The mechanical implementation phase transformed theoretical concepts into practical engineering solutions, constructing the physical prototype and integrating sensors, valves, and control systems. Engineering transitioned to physical implementation through CAD modeling and finite element analysis for pressure vessel design. Component selection criteria prioritized industrial-grade reliability, chemical compatibility with brewing environments, and integration capabilities with digital control systems. Prototype fabrication employed iterative testing cycles, validating mechanical integrity under operational pressures before proceeding to sensor integration. 

\textbf{Phase 3 - Digital Twin Implementation (Winter 2025)}:   This phase implemented the software architecture, database, visualization systems, and bidirectional communication protocols that transform the system from a monitoring device into a true DT. Our software engineering processes followed agile methodology with two-week sprints focusing on incremental functionality. Continuous integration pipelines ensured code quality through automated testing of safety-critical functions. The transition from CAD models to Unity visualization required precise geometry conversion protocols maintaining dimensional accuracy for digital-physical correspondence.

\paragraph{Physical Twin Implementation}

The physical twin integrates key hardware components for data collection and operational control. The fermentation tank serves as the connection point through sample circulation ports and a sealed mounting point for the electronic hydrometer. The sampling chamber comprises a precision-machined stainless steel chamber rated for 15 bar pressures with 10 bar operational limits, incorporating electronically controlled valves, nitrogen injection for pressure control, and sensor ports with sanitary fittings.

The sensor array has an optical dissolved oxygen sensor ($0$-$100$ mg/L range, $\pm0.05$ mg/L accuracy), industrial pH sensor ($0$-$14$ pH range, $\pm0.002$ pH accuracy with temperature compensation), four-electrode conductivity sensor ($1$-$100,000$ µS/cm range), platinum RTD temperature sensor ($\pm0.01\degree$C accuracy across $0$-$30\degree$C range), and pressure transducer (0-7 bar measurement with $\pm0.006$ bar accuracy). The iSpindle electronic hydrometer operates autonomously, providing wireless floating measurement of specific gravity with $\pm0.001$ accuracy and Wi-Fi connectivity for direct data transmission.

\paragraph{Digital Twin Services}
The DT implementation employs a service-oriented architecture comprising five core services: 1) The \textit{Visualization Service} provides real-time dashboards with current parameter values and trends, historical data visualization, and Unity-based 3D representation with animated state visualization\footnote{See demo at \url{https://youtu.be/-EGl8y5Qlus}}. Unity was selected over traditional simulation frameworks (e.g., DEVS) for its real-time rendering capabilities and native WebGL support, enabling browser-based access without specialized software. While DEVS excels at discrete-event simulation, Unity's physics engine provides intuitive 3D interaction for non-technical brewery staff.

%Future iterations will incorporate machine learning capabilities into the Analysis Service, potentially using LSTM networks for fermentation trajectory prediction and anomaly detection based on the accumulated sensor data patterns. 

2) The \textit{Analysis Service} processes brewing data through trend analysis, comparison with theoretical models, and calculation of derived metrics. 3) The \textit{Safety Monitoring Service} performs real-time pressure threshold monitoring and enforcement of physical presence requirements for critical operations.

4) The \textit{Control Service} enables bidirectional interaction through parameter adjustment capabilities for non-critical settings and safe command validation. 5) The \textit{Database Service} provides persistent storage through structured data relationships and time-series optimizations, ensuring brewing knowledge capture and preservation.

\paragraph{Sampling Process}

The fermentation monitoring system follows a cyclic sampling process to collect measurements from the brewing tank without disrupting fermentation. This automated process ensures consistent data collection while maintaining sterile conditions and equipment safety.

\begin{figure}[h!]
\centering

\includegraphics[width=0.485\textwidth]{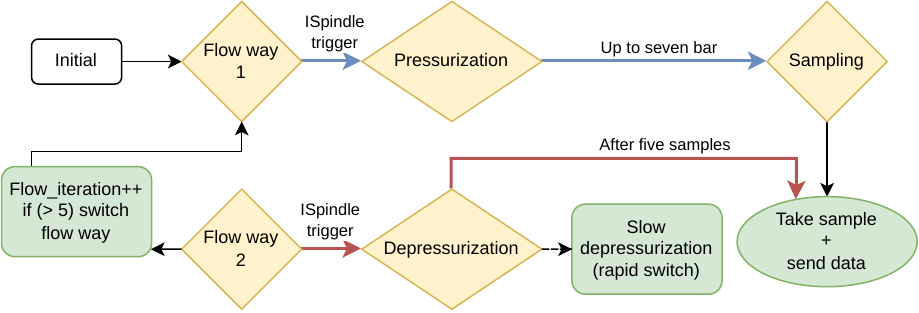}

\caption{Sample collection and data transmission process.}% illustrating the dual-path fermentation monitoring system.}
\label{fig:sample-collection-flowchart}
\end{figure}

Fig.~\ref{fig:sample-collection-flowchart} illustrates the complete sampling cycle implemented as a finite state machine. The process begins in the \textit{Initial} state, triggered by the iSpindle hydrometer. The system then activates \textit{FlowWay1}, which takes a sample from the fermentation tank into the measurement chamber. Once filled, the \textit{Pressurization} state injects nitrogen gas (which does not alter the beer's taste) to achieve 7 bar pressure—necessary for accurate dissolved oxygen measurements. In the \textit{Sampling} state, all sensors' data is immediately transmitted to the DT. 

After measurements are complete, the critical \textit{Depressurization} phase begins, using rapid valve switching to slowly release pressure and prevent equipment damage. Finally, \textit{FlowWay2} returns the sample to the fermentation tank. The system repeats this cycle five times before switching flow directions to ensure representative sampling. Each state transition includes safety checks for pressure thresholds and timing constraints, ensuring predictable operation while preventing hazardous conditions.

% Figure for single column width

\paragraph{Safety-Critical Systems Engineering}
The system implements three distinct safety levels governing access and control permissions. Standard operations including non-critical parameter adjustments can be performed remotely through the DT interface. Limited operations require confirmation at both digital and physical interfaces, creating dual-verification protocols. Critical operations with significant safety implications are restricted to require physical presence at the brewing site.

Safety implementation includes software-based pressure monitoring with automatic shutdown triggers, hardware pressure relief valves providing mechanical failsafe functionality, and watchdog timers detecting controller failures with safe state transitions during communication timeouts.

\paragraph{Safety Requirements Specification}
The system defines specific safety requirements for the pressurized sampling chamber operation. Safety thresholds include a maximum operational pressure of 7 bar with automatic shutdown if pressure exceeds 8 bar, mechanical pressure relief valve set at 10 bar, and software monitoring with 500ms response time for pressure anomalies. While formal safety certification was beyond the project scope, safety validation was performed through systematic testing of failure scenarios including power loss during pressurization, valve malfunction simulations, and sensor failure modes.

%While formal safety certification was beyond the project scope, we conducted hazard analysis identifying key failure points and implemented these safety mechanisms  to ensure safe operation during our testing phase.

%(software monitoring, hardware relief valve, and manual emergency stop)

\paragraph{Digital Twin Architecture}

The DT has software components that collectively process, store, analyze, and visualize data from the physical system. The application server uses Python with Flask, providing RESTful API endpoints, implements publish-subscribe patterns for real-time data distribution, has a safety monitor module, and coordinates bidirectional communication between system components. 

The PostgreSQL database employs specialized schema design with time-series optimization for efficient storage and retrieval of measurement data with automatic aggregation for historical trend analysis. The Unity visualization provides an interactive environment with detailed 3D models, real-time animation, current sensor readings with color-coded status indicators, and interfaces for adjusting non-critical parameters.

\begin{figure*}[tbh]
\centering
\includegraphics[width=0.7\linewidth]{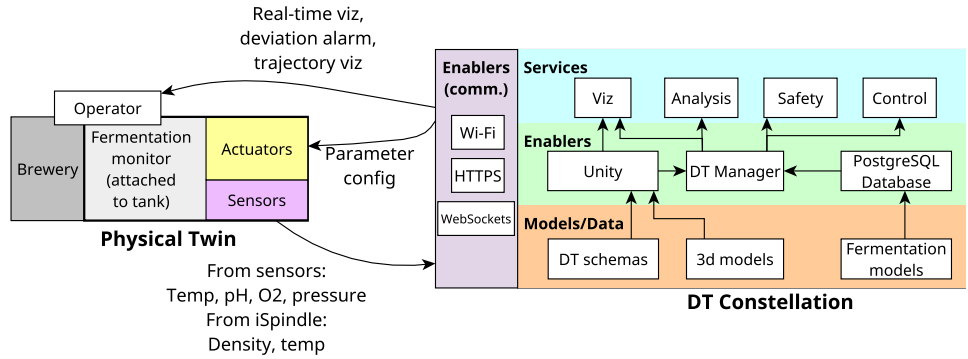}
\caption{Digital Twin constellation architecture~\cite{gil2024toward}.}% showing precise component boundaries, data flow protocols (WiFi, HTTPS, WebSockets), and service interdependencies. Solid arrows indicate real-time data streams; dashed arrows represent control commands.}
\label{fig:dt-constellation}
\end{figure*}

Fig.~\ref{fig:dt-constellation} illustrates how sensor data transforms into actionable insights in the DT constellation~\cite{gil2024toward}. Raw measurements from physical sensors (temperature, pH, conductivity, oxygen, pressure) and the iSpindle hydrometer (density, temperature) flow through communication enablers (WiFi, HTTPS, Web-sockets) to the DT Manager. The \textit{Analysis} service processes this data against previous fermentation data to detect anomalies and predict fermentation completion. When deviations occur, the \textit{Safety} service triggers alarms while the \textit{Visualization} service displays parameter trajectories on the operator's dashboard. Based on these insights, operators can take corrective actions—adjusting sampling frequencies or modifying process parameters—which flow back through the \textit{Control} service to the physical system. This bidirectional data flow enables proactive fermentation management, transforming reactive brewing practices into predictive quality control.

	\section{Engineering Specifics}

% This section describes the systematic engineering approaches throughout the development of the brewing monitoring system, including system integration strategies, calibration procedures, testing protocols, and continuous process optimization techniques for reliable and accurate performance.

\paragraph{System Integration Engineering}

The complexity of integrating mechanical, electrical, and software subsystems necessitated structured engineering approaches. We developed custom Arduino libraries that abstract hardware complexity into logical operations, implementing hardware abstraction layers (HAL) that enable software modifications without requiring hardware changes.

Communication protocol selection was driven by reliability requirements and latency constraints. We implemented message queuing for asynchronous operations to prevent data loss during network interruptions and developed heartbeat mechanisms to ensure continuous system health monitoring across all communication channels. The modular architecture features interchangeable sensor modules with standardized software interfaces, supported by configuration files that enable system adaptation without code modifications.

\paragraph{Calibration and Measurement Engineering}

Accurate measurement in a pressurized environment required calibration procedures for the unique challenges of dynamic pressure conditions. Each sensor undergoes three-point calibration at atmospheric pressure, mid-range operational pressure (3.5 bar), and maximum operational pressure (7 bar) to account for pressure-induced measurement deviations. Cross-validation between sensors ensures consistency, with pH and conductivity sensors requiring additional temperature compensation.

The measurement sequence begins with a 30-second measurement stabilization period for pressure equilibration. Sensors activate collecting 10 measurements per sensor. Statistical averaging with confidence interval calculation precedes the application of pressure algorithms, followed by final validation against expected parameter ranges before data transmission.

\paragraph{Testing and Validation Engineering}

Testing validated system performance across operational and edge-case scenarios. Component-level testing included individual sensor validation under controlled conditions simulating fermentation environments, and pressure cycling tests exceeding 1000 cycles to confirm mechanical integrity of seals and fittings.

Integration testing progressed from sensor-Arduino communication validation through network reliability testing. System-level validation was complete testing with water before introducing fermentation media, failure mode testing including power losses, network disconnections, and sensor malfunctions, and performance benchmarking establishing baseline metrics for response times and data throughput. Safety system verification independently validated each safety mechanism including pressure relief valves, software pressure monitoring, and emergency stop functionality. Fault injection testing confirmed appropriate system responses to component failures.

\paragraph{Process Optimization Engineering}

Continuous improvement methodology drove refinements throughout the development cycle. System metrics include sensor response times, data transmission latency, and visualization update rates. Resource utilization improvements included memory footprint optimization for the embedded Arduino implementation, conserving 40\% of available RAM. Reliability engineering implemented redundant data pathways preventing single points of failure, developed graceful degradation modes maintaining core functionality during partial system failures, and created automated recovery for minimizing manual interventions.
	\section{Implementation Challenges and Solutions}

% This section details three cross-cutting challenges and how we addressed them in our project.

\paragraph{Safety and Physical System Challenges}

Implementation of the pressurized fermentation monitoring system presented significant safety engineering challenges. Early prototype testing revealed leakage issues at connection points during pressure cycling, resolved through redesign of connection fittings with improved sealing materials and implementation of gradual pressure ramping. Sensor reliability under pressure conditions was addressed through sourcing industrial-grade sensors and developing pressure-specific calibration protocols.

The brewery environment introduced electrical noise affecting sensor readings, mitigated with appropriate shielding and implementation of signal filtering at both hardware and software levels. Water damage incidents during prototype testing were addressed through redesign of liquid pathway sections to include overflow protection chambers and implementation of waterproof enclosures for vulnerable electronic components.

\paragraph{Software and Communication Challenges}

Maintaining accurate synchronization between physical and DTs required addressing network latency issues, component clock drift, and data buffering during connection interruptions. Solutions included  timestamp reconciliation, connection health monitoring, and data buffering with catch-up mechanisms.

The Unity visualization system initially faced performance issues with real-time updates, resolved through optimization of 3D models for web performance and implementation of selective update strategies. Database performance with large sensor data volumes was addressed through time-series optimizations in PostgreSQL and development of data retention policies with automatic aggregation.

\paragraph{Interdisciplinary Integration}

The project's interdisciplinary nature spanning brewing science, electrical engineering, software development, mechanical engineering, and process control created unique integration challenges. Knowledge domain integration was addressed through regular cross-disciplinary knowledge-sharing sessions and creation of a common vocabulary for system components. Testing environment limitations were addressed through simulator development for fermentation processes and creation of test scenarios.

\section{Results and Discussion}

% Here we discuss the performance, impact, and limitations of our implemented system.

\paragraph{System Performance}

The implemented DT achieves bidirectional communication between physical and digital components, enabling real-time monitoring of critical fermentation parameters with data collection intervals configurable from 5 seconds to 15 minutes. The safety-critical control systems have demonstrated reliable operation through over 500 pressure cycles without incident, validating our multi-layered safety architecture. The Unity visualization system provides responsive 3D representation with less than 100ms latency for parameter updates, while the database system manages over 1 million data points per batch. 

The DT constellation~\cite{gil2024toward} (seen in Figure~\ref{fig:dt-constellation}) has proven effective as a framework for facilitating interdisciplinary collaboration, with documented improvements in cross-team communication and system understanding. Specifically, the four-layer organizational framework (physical environment, enablers, services, and models/data) created a shared visual language that reduced technical miscommunication between mechanical, and software teams eliminating ambiguity about component relationships and data flow directions.

Prior to using the constellation model, cross-disciplinary discussions often resulted in terminology conflicts and conceptual misalignment, particularly when brewing technicians described fermentation requirements to software teams or when mechanical pressure system modifications required corresponding updates to both the Arduino control logic and the Unity visualization components. %when mechanical teams communicated pressure system constraints to the control systems team.

%The constellation diagram   It also facilitated more effective resource allocation, as team members could clearly identify which layer their expertise contributed to and understand dependencies with other disciplines. This architectural clarity proved especially valuable during the critical integration phases when mechanical pressure system modifications required corresponding updates to both the Arduino control logic and the Unity visualization components.

\paragraph{Value Delivery to Stakeholders}

The DT has delivered quantifiable value to Polybroue operations since first deployment. Brewing technicians have experienced significant operational improvements, with manual sampling time reduced from 45 minutes per batch per day to 4 minutes, representing a 91\% reduction in labor requirements for fermentation monitoring. The system has issued 23 automated alerts across monitored batches. Data collection frequency increased from bi-daily manual measurements to continuous 5-second intervals, providing 17,280 data points per day versus the previous 8 manual readings. This dramatic increase in data granularity has revealed fermentation patterns that were previously invisible to manual monitoring approaches. From an educational perspective, the system serves as a demonstration platform for students across engineering courses. This practical application has proven invaluable for bridging the gap between theoretical knowledge and real-world implementation.

\paragraph{Limitations}

The current implementation presents limitations that constrain broader deployment. The system's cost is high due to industrial-grade sensor requirements and custom hardware integration, limiting accessibility for smaller brewing operations. Scalability challenges emerge when monitoring multiple fermentation vessels simultaneously, as network bandwidth and database performance requirements increase substantially. The system requires specialized technical knowledge for maintenance and troubleshooting, requiring trained personnel. Additionally, the current sensor suite provides limited coverage of certain fermentation parameters such as foam dynamics and yeast viability, requiring supplementary manual testing for comprehensive process monitoring.

\paragraph{Lessons Learned}
The development process has yielded important insights for future DT implementations. The safety-first design philosophy proved critical, with early integration of safety considerations enabling architectural decisions that prevented potential hazards. Progressive implementation strategy evolving from passive monitoring to active control provided validation of foundational components before introducing bidirectional control complexity. Simulation-driven development enabled rapid iteration without risking physical equipment, while rigorous documentation discipline ensured information consistency across domains.

	\section{Related work}

\paragraph*{Academic literature}

Pogo \textit{et al.} implemented a virtual environment for the mashing and brewing of craft beer process~\cite{Pogo2023Virtual}, combining a Unity 3D environment, controller code on a Raspberry PI, and detailed simulations. However, their solution is not connected in real-time to the physical systems. 

The University of Technology Sydney developed a DT for a microbrewery connected to TU Dortmund University in Germany via a Nokia private 5G network, enabling process optimization through cloud-based DT analytics~\cite{deuse2024establishing}. Des-wosu \textit{et al.} examine the use of DTs (as simulation) to track beer mash temperature and to perform predictive maintenance~\cite{azubuike2024}.

Lee \textit{et al.} evaluated the applicability of open-source DT frameworks by applying them to a small fermentation monitoring case~\cite{lee2024evaluating}. They focus on the software development aspect, and do not report overall DT engineering challenges. Wang \textit{et al.} demonstrate DTs for predictive control and forecasting in kombucha fermentation, including hybrid data-driven and mechanistic models for process flexibility and safety~\cite{wang2025}.

%DT technology has gained significant traction in the brewing industry, with implementations to fermentation monitoring and brewery optimization. For example, 

\paragraph*{Commercial solutions}

Anheuser-Busch InBev has implemented brewing and supply chain DTs that enable brewers to adjust inputs based on active conditions and automatically compensate for production bottlenecks, such as when vats are full~\cite{siemens2022}. The ABB Ability\texttrademark~BeerMaker intelligent process control solution includes a DT component for simulating new recipes and testing process modifications in real-time~\cite{siemens2022}. 

Carlsberg's Beer Fingerprinting Project, completed in collaboration with Microsoft and Danish universities, utilized AI and digital twin concepts to analyze hundreds of different beers and predict flavor profiles\cite{siemens2022}.

Solutions such as Precision Fermentation's BrewMonitor System 2.0 provide real-time, sensor-driven fermentation monitoring and analytics, supporting process transparency, consistency, and proactive risk management in breweries~\cite{precision2025brewiq}. 

While these implementations demonstrate the growing adoption of DT technology in brewing applications, most focus on recipe optimization, supply chain management, or general process monitoring. Our work distinguishes itself by specifically addressing safety-critical pressurized fermentation monitoring with bidirectional control capabilities, implementing comprehensive safety protocols for Type 2 Interactive Digital Twin functionality in hazardous brewing environments.%~\cite{anvil2025}.
	
\section{Conclusion}

% \subsection{Summary of Contributions}

This paper presents the successful engineering and implementation of a DT solution for beer fermentation monitoring that advances the state of practice in process engineering applications. The developed system achieves the critical transformation from a digital shadow to an interactive Type 2 Digital Twin~\cite{HVMCatapult2018} through  bidirectional communication protocols, safety-critical control mechanisms, and real-time visualization.

The implementation of a service-oriented architecture comprising visualization, analysis, safety monitoring, control, and database services provides a robust foundation for scalable process monitoring applications. The development of comprehensive Unity-based visualization delivers intuitive three-dimensional representation of fermentation processes, bridging the gap between technical complexity and user accessibility.

The safety-critical systems engineering strategies represent a valuable contribution to the DT engineering community. The successful implementation of multi-layered safety protocols for pressurized operations up to seven bar, combined with sophisticated decompression control mechanisms, demonstrates that interactive DTs can be safely deployed in potentially hazardous industrial environments. The constellation architecture methodology facilitates interdisciplinary collaboration and provides a replicable framework for future DT implementations spanning multiple engineering domains.

\subsection*{Future Work and Research Directions}

Future development directions include expansion to multi-vessel monitoring capabilities and integration of machine learning algorithms for predictive fermentation modeling using the high-resolution datasets generated by the system. The DT framework provides foundation for broader brewing process optimization, including mashing temperature control and packaging line monitoring. The modular architecture enables adaptation for other fermentation industries including distilling, kombucha production, or development of specialized training programs for process control engineering.

%Several promising avenues for future development emerge from this foundational implementation.

The integration of machine learning capabilities represents an immediate opportunity for enhancement. The comprehensive dataset generated by continuous monitoring provides the foundation for developing predictive models that could anticipate fermentation outcomes, detect anomalous conditions, and optimize process parameters automatically. Advanced analytics incorporating time-series forecasting and anomaly detection algorithms could transform the system from reactive monitoring to proactive process optimization.

%Expansion of the DT constellation to encompass the complete brewing operation presents significant opportunities for value creation.
Integration with upstream processes including milling, mashing, and wort production could provide holistic visibility into the entire brewing workflow. Similarly, extension to downstream processes such as conditioning, filtration, and packaging would create a comprehensive digital representation of brewery operations enabling optimization across the complete production chain. 

Research into adaptive control algorithms represents another promising direction. The current implementation enables parameter adjustment within safety constraints, but future development could incorporate autonomous optimization algorithms that continuously adjust process parameters based on real-time fermentation dynamics and historical performance data. The integration of augmented reality interfaces could enhance the visualization capabilities by overlaying DT data onto physical equipment during maintenance and operation activities. This would further bridge the digital and physical representation gap while providing practical utility for brewery personnel.

Finally, the development of standardized DT frameworks for process industries, building upon the constellation architecture demonstrated in this work and the tooling in~\cite{fiter2025DTInsight}, could accelerate adoption of DT technology across manufacturing applications. This standardization effort would require collaboration across academic institutions, technology vendors, and industry stakeholders to establish common protocols and best practices for DT implementation in safety-critical environments.

	\section*{Acknowledgments}
	The authors sincerely thank Dr.~Roozbeh Aghili, the course assistant for the Winter 2025 DT Engineering course. % We extend our heartfelt gratitude to Professor Bentley Oakes for his wonderful guidance and help in writing this article, as well as for his exceptional instruction of the class of 2025.
	We also wish to thank all of Polybroue, especially our colleagues Charles Thiboutot-Lessard, Fabien Portas, and Vincent Paradis for their contributions and support.% throughout this endeavor.
	
	\newpage
	\bibliographystyle{IEEEtran}
	\bibliography{references}

\begin{thebibliography}{10}
\providecommand{\url}[1]{#1}
\csname url@samestyle\endcsname
\providecommand{\newblock}{\relax}
\providecommand{\bibinfo}[2]{#2}
\providecommand{\BIBentrySTDinterwordspacing}{\spaceskip=0pt\relax}
\providecommand{\BIBentryALTinterwordstretchfactor}{4}
\providecommand{\BIBentryALTinterwordspacing}{\spaceskip=\fontdimen2\font plus
\BIBentryALTinterwordstretchfactor\fontdimen3\font minus
  \fontdimen4\font\relax}
\providecommand{\BIBforeignlanguage}[2]{{%
\expandafter\ifx\csname l@#1\endcsname\relax
\typeout{** WARNING: IEEEtran.bst: No hyphenation pattern has been}%
\typeout{** loaded for the language `#1'. Using the pattern for}%
\typeout{** the default language instead.}%
\else
\language=\csname l@#1\endcsname
\fi
#2}}
\providecommand{\BIBdecl}{\relax}
\BIBdecl

\bibitem{Josey2019current}
M.~Josey and D.~L. Maskell, ``Current practices and novel developments in the
  fermentation of beers with non-conventional yeasts,'' \emph{Molecules},
  vol.~24, no.~8, p. 1568, 2019.

\bibitem{Kritzinger2018}
W.~Kritzinger, M.~Karner, G.~Traar, J.~Henjes, and W.~Sihn, ``Digital twin in
  manufacturing: A categorical literature review and classification,''
  \emph{IFAC-PapersOnLine}, vol.~51, no.~11, pp. 1016--1022, 2018.

\bibitem{gil2024toward}
S.~Gil, B.~Oakes, C.~Gomes, M.~Frasheri, and P.~G. Larsen, ``Toward a
  systematic reporting framework for digital twins: a cooperative robotics case
  study,'' \emph{Simulation}, vol. 101, no.~3, pp. 313--339, 2025.

\bibitem{Pogo2023Virtual}
S.~I. Pogo, J.~Granizo, V.~H. Andaluz, and J.~Varela-Aldas, ``Virtual
  environment for the mashing and boiling process in craft beer production,''
  in \emph{Proceedings of 7th International Congress on Information and
  Communication Technology}, X.-S. Yang, S.~Sherratt, N.~Dey, and A.~Joshi,
  Eds.\hskip 1em plus 0.5em minus 0.4em\relax Springer Nature Singapore, 2023,
  pp. 419--435.

\bibitem{deuse2024establishing}
J.~Deuse, R.~W{\"o}stmann, M.~Syberg, N.~West, D.~Wagstyl, and V.~H. Moreno,
  ``Establishing a machine learning and internet of things learning
  infrastructure by operating transnational cyber-physical brewing labs,'' in
  \emph{Conference on Learning Factories}.\hskip 1em plus 0.5em minus
  0.4em\relax Springer, 2024, pp. 171--178.

\bibitem{azubuike2024}
\BIBentryALTinterwordspacing
G.~D. Azubuike, D.~O. Aikhuele, and H.~U. Nwosu, ``Digital twin framework for
  process optimization and predictive maintenance in brewery operations,''
  \emph{Uniport Journal of Engineering \& Scientific Research}, vol.~8, no.~2,
  pp. 337--345, 2024. [Online]. Available:
  \url{https://www.ujesr.org/images/vol82/Article-32.pdf}
\BIBentrySTDinterwordspacing

\bibitem{lee2024evaluating}
A.~Lee, D.~A.~M. Negrin, and L.~Cleophas, ``Evaluating open-source tools for
  heterogeneous model-based digital twin development: A microbrewery case
  study,'' in \emph{2024 Annual Modeling and Simulation Conference
  (ANNSIM)}.\hskip 1em plus 0.5em minus 0.4em\relax IEEE, 2024, pp. 1--13.

\bibitem{wang2025}
\BIBentryALTinterwordspacing
Y.~Wang \emph{et~al.}, ``Digital twin for predicting and controlling food
  fermentation,'' \emph{Computers \& Industrial Engineering}, 2025. [Online].
  Available:
  \url{https://www.sciencedirect.com/science/article/abs/pii/S0260877425000020}
\BIBentrySTDinterwordspacing

\bibitem{siemens2022}
\BIBentryALTinterwordspacing
{Siemens AG}. (2022) Digitalized solutions for the brewery industry. [Online].
  Available:
  \url{https://cdn.worldbeercup.org/wp-content/uploads/2022/12/07104612/vrfb-b10012-00-7600-brewery-industry-en.pdf}
\BIBentrySTDinterwordspacing

\bibitem{precision2025brewiq}
\BIBentryALTinterwordspacing
{Precision Fermentation, Inc.}, ``{BrewIQ: Real-time, comprehensive
  fermentation monitoring},'' 2025, accessed: 2025-07-04. [Online]. Available:
  \url{https://www.precisionfermentation.com/brewiq-system/}
\BIBentrySTDinterwordspacing

\bibitem{HVMCatapult2018}
{High Value Manufacturing Catapult Visualisation and VR Forum}, ``Feasibility
  of an immersive digital twin: The definition of a digital twin and
  discussions around the benefit of immersion,'' High Value Manufacturing
  Catapult, Shirley, UK, Tech. Rep., Sep. 2018, sponsored by Innovate UK;
  published September 2018.

\bibitem{fiter2025DTInsight}
K.~Fiter, L.~Malassign\'{e}-Onfroy, and B.~Oakes, ``{DTInsight}: A tool for
  explicit, interactive, and continuous digital twin reporting,'' in
  \emph{Proceedings of the 28th ACM/IEEE International Conference on Model
  Driven Engineering Languages and Systems: Companion Proceedings}, 2025.

\end{thebibliography}
\end{document}